\documentclass[12pt,letterpaper]{article}
\usepackage{natbib,epsfig,graphicx,setspace}
\usepackage{amsmath,amsthm,amssymb}
\bibpunct{(}{)}{;}{a}{,}{,}

\newcommand{\csection}[1]
    {\begin{center}
        \stepcounter{section}
        {\bf\large\arabic{section}. #1}
    \end{center}
    \vspace{-0.15 cm}
}

\newcommand{\scsection}[1]
    {\begin{center}
        {\bf\large #1}
    \end{center}
    \vspace{-0.15 cm}
}

\newcommand{\csubsection}[1]{
\vspace{-0.25 cm}
\begin{center}
\stepcounter{subsection} {\it\arabic{section}.\arabic{subsection}.
#1}
\end{center}
\vspace{-0.25 cm} }

\def\bg{\begin{figure}[tpbh]\begin{center}}
\def\eg{\end{center}\end{figure}}

\numberwithin{equation}{section} 

\newcommand{\wln}{\xrightarrow{D}}

\newcommand{\e}{\text{E}}
\newcommand{\phih}{\phi_{h_2}}
\newcommand{\var}{\text{Var}}
\newcommand{\cov}{\text{Cov}}

\newcommand{\tr}{\text{tr}}

\newcommand{\beqn}{\begin{eqnarray}}
\newcommand{\eeqn}{\end{eqnarray}}
\newcommand{\beq}{\begin{equation}}
\newcommand{\eeq}{\end{equation}}

\long\def\symbolfootnote[#1]#2{\begingroup\def\thefootnote{\fnsymbol{footnote}}
\footnote[#1]{#2}\endgroup}

\newcommand{\jasa}{\emph{Journal of the American Statistical Association}}

\newcommand{\ann}{\emph{Annals of Statistics}}

\newtheorem{thm}{Theorem}[section]
\newtheorem{lem}{Lemma}[section]

\def\btheta{{\mbox{\boldmath $\theta$}}}

\def\balpha{{\mbox{\boldmath $\alpha$}}}
\def\bx{\bold{x}}
\def\bX{\bold{X}}

\def\bY{\textbf{Y}}

\def\bW{\textbf{W}}

\def\boldtheta{{\mbox{\boldmath $\theta$}}}

\def\balpha{{\mbox{\boldmath $\alpha$}}}

\def\mode{\text{Mode}}

\begin{document}

\title{Nonparametric and Varying Coefficient Modal Regression}
\author{Weixin Yao \thanks{
Department of Statistics,  University of California, Riverside, California,
U.S.A. E-mail: weixin.yao@ucr.edu. Yao's research is supported by NSF grant DMS-1461677.}  and Sijia Xiang\thanks{School of Mathematics and Statistics, Zhejiang University of Finance and Economics. E-mail:
sjxiang@zufe.edu.cn. Xiang's research is supported by Zhejiang Provincial NSF of China grant LQ16A010002.}}

\date{}
\maketitle{}

\centerline{\sc Summary}  In this article, we propose a new
nonparametric data analysis tool, which we call \emph{nonparametric modal regression}, to investigate the relationship among interested variables based on estimating the mode of the conditional density of a response variable $Y$ given predictors $X$.
The nonparametric modal regression is distinguished from the
conventional nonparametric regression in that, instead of the
conditional average or median, it uses the ``most likely" conditional values to measures the
center. Better prediction performance and
robustness are two important characteristics of nonparametric modal regression
compared to traditional nonparametric mean regression and nonparametric median regression. We propose to use local polynomial regression
to estimate the nonparametric modal regression. The asymptotic
properties of the resulting estimator are investigated.
To broaden the applicability of the nonparametric modal regression to high dimensional data or functional/longitudinal data, we further
develop a nonparametric varying coefficient modal regression. A Monte Carlo
simulation study and an analysis of health care expenditure data demonstrate
some superior performance of the proposed nonparametric modal regression model
to the traditional nonparametric mean regression and nonparametric median regression in terms of the prediction performance.  

\vspace*{.3in}

\noindent {\it Some key words:} EM algorithm; Local polynomial regression; Modal regression; Mode; Robust.


\csection{\sc Introduction}
Suppose $\{(x_i,
y_i),i=1,\ldots,n\}$ is a random sample, where $x_i$ is a $p$-dimensional column
vector, and $f(y\mid x)$ is the conditional density function
of $Y$ given $x_i$. For the conventional regression models, the mean of $f(y\mid x)$ is usually used to investigate the relationship between $Y$ and $X$. When the distribution is highly skewed, it is well known that the mode provides a more meaningful location estimator than the mean. Several authors have made efforts to identify the modes of population distributions.
See, for example, \citet{Scott92, FF99, CM99,
HMZ04, RL05, yao09}. Recently, Lee (1989, 1993), Lee and Kim (1998), Kemp and Santos Silva (2012), and Yao and Li (2014) successfully applied the mode idea to linear regression and proposed the linear modal regression which assumes that the mode of $f(y\mid x)$, denoted by $\mode(y\mid x)$, is a linear function of $x$. Better prediction performance and robustness are two important characteristics of modal regression compared to the traditional mean regressions. Please see Kemp and Santos Silva (2012) and Yao and Li (2014) for more discussions about the advantage of modal regression as a promising alternative regression tool to traditional regression models. However, in practice, the strong parametric assumption about $\mode(y\mid x)$ might not hold and thus the corresponding inference might be misleading. Therefore, it is desirable to develop some estimation procedure to relax the parametric model assumption about $\mode(y\mid x)$.

In this article, we propose a nonparametric modal regression model
that aims to estimate the \emph{mode} of $f(y\mid x)$ for any
given $x$ without assuming any parametric model. Instead of the conditional average
used by the traditional regression methods, modal regression uses the ``most
probable" conditional values to measure the center. We propose to use
local polynomial regression to estimate the nonparametric modal
regression. Sampling properties of the proposed estimates are systematically studied. A modal expectation-maximization (EM) algorithm is also developed for the proposed models.
To broaden the applicability of the nonparametric modal regression, we further develop a nonparametric varying coefficient modal regression. A Monte Carlo simulation study and an analysis of health care expenditure data demonstrate some superior performance of the proposed nonparametric modal regression to the traditional nonparametric mean regression.

The rest of this article is organized as follows. In Section 2, we
introduce the new nonparametric modal regression model and the
estimation procedure based on local polynomial regression. The
asymptotic properties of the resulting estimator are also
provided. In Section 3, we propose a nonparametric varying coefficient modal regression. A Monte Carlo simulation study and a real data application are conducted in Section 4 to illustrate the proposed models. We conclude our article by some discussions in Section 5. 

 \csection{\sc Nonparametric Modal Regression}
In this section, we will introduce the nonparametric modal
regression model and the estimation procedure based on local polynomial regression. An EM type algorithm
is proposed to estimate the unknown modal parameters. In
addition, we will also study the asymptotic properties of the
proposed estimator.

\csubsection{Model introduction}
Suppose that  $(x_1,y_1),\ldots,(x_n,y_n)$ are an independent and
identically distributed random sample from $f(x,y)$. The modal
regression is defined as
\begin{equation}
m(x)=\mode(y\mid x)=\arg\max_yf(y\mid x), \label{nonmod}
\end{equation}
where $m(\cdot)$ is an unknown nonparametric smoothing function to
be estimated. For simplicity of explanation, we assume that $x$ is
a scalar but the proposed model can be extended to the multivariate
predictor $x$. However, such extension is less desirable due to
the ``curse of dimensionality".

Let $\epsilon=y-m(x)$. Denote by $g(\epsilon\mid x)$ the conditional
density of $\epsilon$ given $X=x$. Based on the model assumption
(\ref{nonmod}), one can know that
$g(\epsilon\mid x)$ is maximized at $0$ for any $x$. 
If $g(\epsilon\mid x)$ is symmetric about $0$, then $m(x)$ is the
same as the conventional regression function $E(Y\mid X=x)$. In
this article, we propose an estimation procedure for the
nonparametric modal regression $m(x)$.

Since $f(y\mid x)=f(x,y)/f(x)$, finding the mode of $f(y\mid x)$ is
equivalent to finding the mode of $f(x,y)$ with $x$ fixed . Suppose
$f(x,y)$ is estimated by the kernel density estimator, i.e.,
\[\hat{f}(x,y)=\frac{1}{n}\sum_{i=1}^nK_{h_1}(x_i-x)\phi_{h_2}(y_i-y),\]
where $K_h(x)=h^{-1}K(x/h)$ and $\phi_{h}(t)=h^{-1}\phi(t/h)$ are
the symmetric kernel functions and $(h_1, h_2)$ are the bandwidths.

A natural estimation procedure is to estimate $m(x_0)$ by
\begin{equation}
\label{eq:modreg}
\hat{m}(x_0)=\arg\max_y\frac{1}{n}\sum_{i=1}^nK_{h_1}(x_i-x_0)\phi_{h_2}(y_i-y)\:.
\end{equation}
Note that the modal regression (\ref{eq:modreg}) only uses one term
(the intercept) for the conditional mode, like the local constant
estimator \citep{nadaraya64, watson64}.  
It is known that the local linear estimator is superior to the local
constant one (Fan and Gijbels, 1996), and so we may want to extend the idea of
local constant modal regression (\ref{eq:modreg}) to the local
linear case, or more generally, the local polynomial case.  For
$x$ in a neighborhood of $x_0$, we approximate
\[
m(x)\approx\sum_{j=0}^p
\frac{m^{(j)}(x_0)}{j!}(x-x_0)^j\equiv\sum_{j=0}^p\beta_j(x-x_0)^j,
\]
where $\beta_j=m^{(j)}(x_0)/j!$. Our local polynomial modal
regression (LPMR) estimation procedure is to \emph{maximize} over
$\boldtheta=(\beta_0, \ldots,\beta_p)$
\begin{equation}
\ell(\btheta)\equiv\frac{1}{n}\sum_{i=1}^nK_{h_1}(x_i-x_0)\phi_{h_2}\left(y_i-\sum_{j=0}^p\beta_j(x_i-x_0)^j\right).
\label{lpmod}
\end{equation}
For ease of computation, we use the standard normal density for
$\phi(t)$  throughout this artical. See (\ref{mstep}) below.
Denote the maximizer of $\ell(\btheta)$ as
$\hat{\btheta}=(\hat{\beta}_0,\cdots, \hat{\beta}_p)$. Then the
estimator of the $v$-th derivative of $m(x)$, $m^{(v)}(x)$, will
be $\hat{m}_v(x_0)=v!\hat{\beta}_v,\text{ for } v=0,\cdots,p.$
Specifically, when $p=1$ and $v=0$, we refer to this method as the
local linear modal regression (LLMR).

Note that Yao et al. (2012) used an objective function similar to (\ref{lpmod}) to provide an adaptive robust nonparametric regression estimate. However, the model setting and the assumptions on tuning parameters in this article are completely different from theirs. Although Yao et al. (2012) also named their method modal regression, similar to the traditional robust regression, they assume that the error distribution is symmetric about 0 in order to get a consistent estimate. Therefore, Yao et al. (2012) still focused on mean regression, even though they motivated their estimation procedure from a modal regression point of view. The nonparametric modal regression we proposed in this article, however, allows the error distribution to be skewed or even depend on $x$ and is truly targeting the conditional mode of $f(y\mid x)$. In addition, in Yao et al. (2012), $h_2$ is a fixed value and does not depend on $n$. In this article, we assume that $h_2$ goes to 0 in order to get the consistent modal regression estimate under very mild assumption of the error distribution. Moreover, we show in Section 2.3 that the asymptotic results, such as convergence rates, of the proposed estimate are completely different from Yao et al. (2012).

\csubsection{Computation algorithm}
Note that (\ref{lpmod}) does not have an explicit solution. Similar to Yao et al. (2012), we can use an EM algorithm to maximize
(\ref{lpmod}) since it has a mixture type form. For easy reference, we also describe the algorithm below. 

Let $\btheta^{(0)}$ be the initial
value. Starting with $k=0$:
\begin{description}
\item{\em E-Step}: Update $\pi(j\mid \btheta^{(k)})$
\begin{align*}
 &\pi(j\mid \boldtheta^{(k)})=\dfrac{K_{h_1}(x_j-x_0)\phi_{h_2}
 \left\{y_j-\sum_{l=0}^p\beta_l^{(k)}(x_j-x_0)^l\right\}}
 {\sum\limits_{i=1}^n\left[K_{h_1}(x_i-x_0)\phi_{h_2}\left\{y_i-
 \sum_{l=0}^p\beta_l^{(k)}(x_i-x_0)^l\right\}\right]}\:,j=1,\ldots,n.
 \end{align*}
 \item{\em M-Step}: Update $\btheta^{(k+1)}$
\begin{eqnarray}
 \boldtheta^{(k+1)}&=&\arg\max\limits_{\boldtheta}\sum\limits_{j=1}^n\left[
 \pi(j\mid \boldtheta^{(k)})\log
\phi_{h_2}\left\{y_j-\sum_{l=0}^p\beta_l(x_j-x_0)^l\right\}\right]\nonumber\\
&=&(\bX^TW_k\bX)^{-1}\bX^T\bW_k\bY,\label{mstep}
\end{eqnarray}
since $\phi(\cdot)$ is the density function of a standard normal
distribution. Here $\bX=(\bx_1^*,\ldots,\bx_n^*)^T$ with
$\bx_i^*=\{1,x_i-x_0,\cdots,(x_i-x_0)^p \}^T,$ $\bW_k$ is an $n\times
n$ diagonal matrix with diagonal elements $\pi(j\mid
\boldtheta^{(k)})$s, and $\bY=(y_1,\ldots,y_n)^T$.
\end{description}
Similar to the usual EM algorithm, the value the algorithm converged to might rely on the starting values, and it is not certain that the algorithm converges to the global optimum. Thus, initiating the algorithm from different starting values and then choosing the best local optimal solution is vital.


\csubsection{Theoretical properties}
\label{sec:theory}
We first establish the convergence rate of the LPMR estimator in
the following theorem, whose proof can be found in the Appendix.
\begin{thm}
\label{modconsis} Under the regularity conditions (A1)|(A4) in the
Appendix, if the bandwidths $h_1$ and $h_2$ go to 0 such that
 $nh_1h_2^5\rightarrow\infty$ and $h_1^{p+1}/h_2\rightarrow 0$, there exists a consistent local maximizer
$\hat{\boldtheta}$ of (\ref{lpmod}) such that
\[
\left|h_1^v\left\{\hat{m}_v(x_0)-m^{(v)}(x_0)\right\}\right|=
O_p\left\{(nh_1h_2^3)^{-1/2}+h_1^{p+1}+h_2^2\right\},\;v=0,1,\ldots,p,
\]
where $\hat{m}_v(x_0)=v!\hat{\beta}_v$ is the estimate of
$m^{(v)}(x_0)$ and $m^{(v)}(x_0)$ is the $v^{\text{th}}$
derivative of $m(x)$ at $x_0$.
\end{thm}

The proof of Theorem \ref{modconsis} is given in the Appendix. To
derive the asymptotic bias and variance of the LPMR estimator, we
need the following notations. The moments of $K$ and $K^2$ are
denoted respectively by
\[\mu_j=\int t^jK(t)dt \qquad \text{and}\qquad \nu_j=\int
t^jK^2(t)dt.\] Let $S$, $\tilde{S}$, and $S^*$  be $(p+1)\times
(p+1)$ matrix with $(j,l)$-element $\mu_{j+l-2}$, $\mu_{j+l-1}$,
and $\nu_{j+l-2}$, respectively, and $c_p$, $\tilde{c}_p$, and
$c_p^*$ be $p\times 1$ vector with $j$-th element $\mu_{p+j}$,
$\mu_{p+j+1}$, and $\mu_{j-1}$, respectively. Furthermore, let
$e_{v+1}=(0,\ldots,0,1,0,\ldots,0)^T$ be a $p\times 1$ vector with 1
in the $(v+1)^{th}$ position.

\begin{thm} \label{modasybv}
Under the regularity conditions (A1)|(A4) in the Appendix, if the bandwidths $h_1$ and $h_2$ go to 0 such that $nh_1^3h_2^5\rightarrow \infty$ and $h_1^{p+1}/h_2\rightarrow 0$, the
asymptotic variance of $\hat{m}_v(x_0)$, given in Theorem
\ref{modconsis}, is
\begin{equation*}
\var\{\hat{m}_v(x_0)\}=e^T_{v+1}S^{-1}S^*S^{-1}e_{v+1}\frac{v!^2g(0\mid
x_0)\tilde{\nu}}{f(x_0)g''(0\mid
x_0)^2nh_1^{1+2v}h_2^3}\left\{1+o_p(1)\right\}, \label{eq:asyvar}
\end{equation*}
where $\tilde{\nu}=\int t^2\phi^2(t)dt.$ The asymptotic bias of
$\hat{m}_v(x_0)$, denoted by $b_v(x_0)$, for $p-v$ odd is given by
\begin{equation*}
b_v(x_0)=e_{v+1}^TS^{-1}\left\{h_1^{p+1-v}\frac{v!}{(p+1)!}m^{(p+1)}(x_0)c_p-\frac{g'''(0\mid
x_0)v!h_2^2}{2g''(0\mid
x_0)h_1^v}c^*_p\right\}\left\{1+o(1)\right\}. \label{eq:asybias1}
\end{equation*}

\noindent Furthermore, the asymptotic bias for $p-v$ even is
\begin{align}
b_v(x_0)=&e_{v+1}^TS^{-1}\left[\tilde{c}_p\frac{h_1^{p+2-v}v!}{(p+2)!}\left\{m^{(p+2)}(x_0)+(p+2)m^{(p+1)}(x_0)
\frac{\Gamma'(x_0)}{\Gamma(x_0)}\right\}\right.\notag\\
&\ \ \left.-\frac{g'''(0\mid x_0)v!h_2^2}{2g''(0\mid
x_0)h_1^v}c^*_p\right]\left\{1+o(1)\right\}, \label{eq:asybias2}
\end{align}
provided that $m^{(p+2)}(\cdot)$ is continuous in a neighborhood of
$x_0$, where
$\Gamma(x)=g''(0\mid x)f(x)$.
\end{thm}

The proof of Theorem~\ref{modasybv} is given in the Appendix.
Similar to the local polynomial regression (LPR), the second term in (\ref{eq:asybias2}) often
creates extra bias and depends on the design density $f(x)$. Thus,
it is preferable to use odd values of $p-v$ in practice. Therefore,
it is consistent with the selection
order of $p$ for the LPR \citep{fan96}. 

\begin{thm}
\label{modnorm} Under the regularity conditions (A1)|(A4) in the
Appendix, if the bandwidths $h_1$ and $h_2$ go to 0 such that
 $nh_1h_2^5\rightarrow\infty$ and $h_1^{p+1}/h_2\rightarrow 0$, the estimate $\hat{m}_v(x_0)$, given in Theorem
\ref{modconsis}, has the following asymptotic distribution
\[
\frac{\hat{m}_v(x_0)-m^{(v)}(x_0)-b_v(x_0)}{\sqrt{\var\{\hat{m}_v(x_0)\}}}
\wln N(0,1).
\]
\end{thm}
The proof of Theorem \ref{modnorm} is given in the Appendix. 
Specially, when $p=1$ and $v=0$, the asymptotic variance of
$\hat{m}(x_0)$ is
\begin{equation*}\var\{\hat{m}(x_0)\}\approx\frac{g(0\mid
x_0)\tilde{\nu}\nu_0}{nh_1h_2^3g''(0\mid
x_0)^2f(x_0)}\left\{1+o_p(1)\right\},
\end{equation*} and the
asymptotic bias is
\begin{equation*}
b(x_0)\approx \frac{1}{2}m''(x_0)\mu_2h_1^{2}-\frac{g'''(0\mid
x_0)h_2^2}{2g''(0\mid x_0)}.
\end{equation*}
To find the global optimal bandwidth, we proposed to minimize the asymptotic weighted mean integrated squared error given by
\[\int\left([\text{Bias}\{\hat{m}(x_0)\mid X\}]^2+\var\{\hat{m}(x_0)\mid X\}\right)w(x)dx =\frac{K}{nh_1h_2^3}+Mh_1^4+Nh_2^4+2Lh_1^2h_2^2,\]
where
\begin{align*}
K=\int\frac{g(0\mid x)\tilde{\nu}\nu_0}{g''(0\mid x)^2f(x)}w(x)dx&, \quad M=\int\left[\frac{1}{2}m''(x)\mu_2\right]^2w(x)dx,\\
N=\int\left[-\frac{g'''(0\mid x)}{2g''(0\mid x)}\right]^2w(x)dx&, \quad L=\int\left[\frac{1}{2}m''(x)\mu_2\right]\left[-\frac{g'''(0\mid
x)}{2g''(0\mid x)}\right]w(x)dx,
\end{align*}
and $w(x)$ is a weight function, such as 1 or the design density $f(x)$.
Therefore, the asymptotic global optimal bandwidth is
\begin{equation}
\hat{h}_1=\left[\frac{3K}{4n\delta^5(L+N\delta^2)}\right]^{1/8}, \hat{h}_2=\delta\hat{h}_1,
\label{bandwidth}
\end{equation}
where $\delta^2=(\sqrt{L^2+3MN}+L)/N$.

 \csection{\sc Nonparametric Varying Coefficient Modal Regression}
Next, we will introduce
how to apply the idea of varying-coefficient models to modal
regression. Varying coefficient models (Hastie and Tibshirani, 1993) have been successfully
applied to high-dimensional nonparametric regression, generalized
linear models, longitudinal and functional data analysis, and
others. Please see, for example, Hoover, Rice, Wu, and Yang (1998); Wu, Chiang, and Hoover (1998); Fan and Zhang (1999, 2000); Chiang, Rice, and Wu (2001); Huang, Wu, and Zhou (2002), for more details.

Given a random sample $\{(\bx_i,u_i,y_i), i=1,...,n\}$, where
$y_i$ is the response variable, $\bx_i$ is a $p-$dimensional
predictor (with first component equals 1), and $u_i$ is a scalar predictor. Suppose $f(y\mid
\bx_i,u_i)$ is the conditional density function of $y$ given $(
\bx_i,u_i)$. The \emph{varying coefficient modal regression }assumes
\begin{equation}
\mode(y\mid\bx_i,u_i)=\max\limits_yf(y\mid
\bx_i,u_i)=\sum_{j=1}^pg_j(u_i)x_{ij}, \label{varcoe}
\end{equation}
where $\bx_i=(x_{i1},\ldots,x_{ip})^T$ and
$\{g_1(u),\ldots,g_p(u)\}^T$ are unknown smooth functions. Note that
nonparametric modal regression (\ref{nonmod}) is a special case of
(\ref{varcoe}) if we take $p=1$ and $x_i=1$. The proposed method can
be easily extended to the case when $U$ is multivariate. However,
the extension to the multivariate $U$ might be practically less
useful due to the ``curse of dimensionality".

We propose to estimate the varying coefficient modal regression by
\emph{maximizing}
\begin{equation}
\ell^*(\btheta)=\sum_{i=1}^nK_{h_1}(u_i-u_0)\phi_{h_2}
\left[y_i-\sum_{j=1}^p\left\{b_j+c_j(u_i-u_0)\right\}x_{ij}\right],
\label{def:vcloss}
\end{equation}
where $\btheta=(b_1,\ldots,b_p,h_1c_1,\ldots,h_1c_p)^T$. 
 For the simplicity of explanation,
here we only consider the local linear approximation based on the
arguments following Theorem \ref{modasybv}. 

We can use an algorithm similar to the EM algorithm proposed in Section 2.2
to maximize (\ref{def:vcloss}). Starting with $k=0$:
\begin{description}
\item{\em E-Step}: Update $\pi(j\mid \btheta^{(k)})$
\begin{align*}
 &\pi(j\mid \btheta^{(k)})=\dfrac{K_{h_1}(u_j-u_0)\phi_{h_2}
\left[
y_j-\sum_{l=1}^p\left\{b_{l}^{(k)}+c_{l}^{(k)}(u_j-u_0)\right\}x_{jl}\right]}
 {\sum\limits_{i=1}^nK_{h_1}(u_i-u_0)\phi_{h_2}\left[
y_i-\sum_{l=1}^p\left\{b_{l}^{(k)}+c_{l}^{(k)}(u_i-u_0)\right\}x_{il}\right]}\:,j=1,\ldots,n.
 \end{align*}
 \item{\em M-Step}: Update $\btheta^{(k+1)}$
\begin{eqnarray}
 \btheta^{(k+1)}&=&\arg\max\limits_{\btheta}\sum\limits_{j=1}^n
 \pi(j\mid \btheta^{(k)})\log
\phi_{h_2}\left[
y_j-\sum_{l=1}^p\left\{b_{l}^{(k)}+c_{l}^{(k)}(u_j-u_0)\right\}x_{jl}\right]\nonumber,
\end{eqnarray}
which has explicit solution since $\phi(\cdot)$ is the Gaussian density.
\end{description}

Denote by $f(u)$ the marginal density of $u$, $q(\epsilon\mid
\bx,u)$ the conditional density of
$\epsilon=y-\sum_{j=1}^pg_j(u)x_j$ given $\bx$ and $u$, and
$q^{(v)}(\epsilon\mid \bx,u)$ the $v$-th derivative of
$q(\epsilon\mid \bx,u)$. Let
\begin{align*}
\alpha_j(u)&= \e\{\bx X_jq^{(2)}(0\mid \bx,u)\mid u\},\quad
\beta(u)=\e\{\bx q^{(3)}(0\mid \bx,u)\mid
u\}\\
\Delta(u)&= \e\{\bx\bx^Tq^{(2)}(0\mid \bx,u)\mid u\},\quad
                       \tilde{\Delta}(u)= \e\{\bx\bx^Tq(0\mid \bx,u)\mid u\}.\label{h}
\end{align*}
Suppose $\hat{\btheta}$ is the maximizer of
(\ref{def:vcloss}), then
$\hat{\textbf{g}}(u_0)=(\hat{b}_1,\ldots,\hat{b}_p)^T$ is the
estimate of $\{g_1(u_0),\ldots,g_p(u_0)\}^T$, and $\hat{\textbf{g}}'(u_0)=(\hat{c}_1,\ldots,\hat{c}_p)^T$ is the
estimate of $\{g'_1(u_0),\ldots,g'_p(u_0)\}^T$.

\begin{thm} \label{modasyvarcoef}
Under the regularity conditions (A5)|(A8) in the Appendix, if the bandwidths $h_1$ and $h_2$ go to 0 such that
 $nh_1^3h_2^5\rightarrow\infty$ and $h_1^{2}/h_2\rightarrow 0$ the
asymptotic bias of $\hat{\textbf{g}}(u_0)$ is given by
\begin{equation}
\text{Bias}\left\{\hat{\textbf{g}}(u_0)\right\}=\frac{1}{2}\Delta^{-1}(u_0)\left\{\mu_2h_1^2\sum_{j=1}^pg_j''(u_0)\alpha_j(u_0)-h_2^2\beta(u_0)\right\}\left\{1+o_p(1)\right\},
\label{varcoeparbias}
\end{equation}
and the asymptotic covariance is
\begin{equation}
\cov\left\{\hat{\textbf{g}}(u_0)\right\}=\frac{\tilde{\nu}\nu_0}{nh_1h_2^3f(u_0)}\Delta^{-1}(u_0)\tilde{\Delta}(u_0)\Delta^{-1}(u_0)\left\{1+o_p(1)\right\},
\label{varcoeparvar}
\end{equation}
where $\mu_j=\int t^jK(t)dt, \nu_j=\int t^jK^2(t)dt,$ and $\tilde{\nu}=\int t^2\phi^2(t)dt$.
\end{thm}

\begin{thm} \label{modasynorm}
Under regularity conditions (A5)|(A8) in the Appendix, if the bandwidths $h_1$ and $h_2$ go to 0 such that
 $nh_1h_2^5\rightarrow\infty$ and $h_1^{2}/h_2\rightarrow 0$, the estimate $\textbf{g}(u_0)$ has the following asymptotic distribution
\[[\cov\{\hat{\textbf{g}}(u_0)\}]^{-1/2}[\hat{\textbf{g}}(u_0)-\textbf{g}_0(u_0)-\text{Bias}\{\hat{\textbf{g}}(u_0)\}]\wln N(0,I),\]
where $\text{Bias}\{\hat{\textbf{g}}(u_0)\}$ is defined in
(\ref{varcoeparbias}) and  $\cov\{\hat{\textbf{g}}(u_0)\}$ is
defined in (\ref{varcoeparvar}).
\end{thm}

The asymptotic  global optimal bandwidth can be found by minimizing the asymptotic weighted mean integrated squared error given by
\begin{align*}
&\int\e\left[\{\hat{\textbf{g}}(u)-\textbf{g}_0(u)\}^TW\{\hat{\textbf{g}}(u)-\textbf{g}_0(u)\}\right]w(u)du\\
\approx&
\int\left[Bias\{\hat{\textbf{g}}(u)\}^TWBias\{\hat{\textbf{g}}(u)\}+\tr\left[Cov\{\hat{\textbf{g}}(u)\}W\right]\right]w(u)du,
\end{align*}
where $W$ is a weight matrix and $w(u)$ is a weight function, such as 1 or the design density for $u$. One
popular choice for $W$ is $[\Delta^{-1}(U)\tilde{\Delta}(U)\Delta^{-1}(U)]^{-1}$, which is proportional to the inverse of the asymptotic
variance of $\hat{\textbf{g}}(u)$.
Based on the asymptotic results of (\ref{varcoeparbias}) and
(\ref{varcoeparvar}), the theoretical global optimal bandwidths are
\begin{equation*}
\hat{h}_1=\left[\frac{3\tilde{K}}{4n\delta^5(\tilde{L}+\tilde{N}\tilde{\delta}^2)}\right]^{1/8}, \quad \hat{h}_2=\tilde{\delta}\hat{h}_1,
\label{bandwidth2}
\end{equation*}
where
\begin{align*}
\tilde{K}&=p\tilde{\nu}\nu_0\int f(u)^{-1}w(u)du,\quad \tilde{L}=-\mu_2\int\sum_{j=1}^p\alpha_j(u)\tilde{\Delta}(u)^{-1}\beta(u)w(u)du\\
\tilde{N}&=\int\beta(u)'\tilde{\Delta}(u)\beta(u)w(u)du,\quad\tilde{\delta}^2=(\sqrt{\tilde{L}^2+3\tilde{M}\tilde{N}}+\tilde{L})/\tilde{N}, \\ \tilde{M}&=\mu_2^2\int\{\sum_{j=1}^pg_j''(u)\alpha_j'(u)\}\tilde{\Delta}^{-1}(u)\{\sum_{j=1}^pg_j''(u)\alpha_j'(u)\}w(u)du.
\end{align*}
We will also investigate how to practically choose
the bandwidth based on the above theoretical results in the simulation study.

\csection{\sc Simulation Study and Application} In this section,
we will use a  Monte Carlo simulation study and a real data application to assess the performance
of the proposed nonparametric modal regression.

To use the proposed two nonparametric modal regression models, we need to select the bandwidths first. Note that the asymptotic global optimal bandwidth formula (\ref{bandwidth}) contains the unknown quantities $m''(x)$ and $g^{(\nu)}(0|x)$, $\nu=0,2,3$, the $\nu$-th derivative of conditional density of $\epsilon$ given $x$, and therefore, is not ready to use. One possible practical way is to apply the plug-in method by replacing the unknown quantities with some estimates. We propose to estimate $m(x)$ by a polynomial function of order three, i.e., $m(x)\approx\tilde{\bx}^T\balpha$, where $\tilde{\bx}=(1,x,x^2,x^3)^T$ and $\balpha=(\alpha_0,\alpha_1,\alpha_2,\alpha_3)^T$. We can then estimate $\epsilon_i$ by $\hat{\epsilon}_i=y_i-\tilde{\bx}_i^T\hat{\balpha}$ and $m''(x)$ by $\hat{m}''(x)=2\hat{\alpha}_2+6\hat{\alpha}_3x$, where $\hat{\balpha}$ is the modal linear regression estimator (Yao and Li, 2013). In our simulation, $\epsilon$ and $x$ are independent. Therefore, $\hat{\epsilon}_i-\hat{m}(x_i)$ has approximate density $g(\cdot)$, and $g^{(\nu)}(0|x)$ can be estimated by
\[\hat{g}^{(\nu)}(0|x)=\frac{1}{h^{\nu+1}}\sum_{i=1}^nK^{(\nu)}\left\{\frac{\hat{\epsilon}_i-\hat{m}(x_i)}{h}\right\},\nu=0,2,3.\]
If $w(x)$ in (\ref{bandwidth}) is equal to the design density $f(x)$, then $K$, $M$, $N$, and $L$ can be estimated by their empirical version:
\begin{align*}
K=\frac{1}{n}\sum_{i=1}^n\frac{\hat{g}(0\mid x_i)\tilde{\nu}\nu_0}{\hat{g}''(0\mid x_i)^2f(x_i)}&, M=\frac{1}{n}\sum_{i=1}^n\left\{\frac{1}{2}\hat{m}''(x_i)\mu_2\right\}^2,\\
N=\frac{1}{n}\sum_{i=1}^n\left\{-\frac{\hat{g}'''(0\mid x_i)}{2\hat{g}''(0\mid x_i)}\right\}^2&, L=\frac{1}{n}\sum_{i=1}^n\left\{\frac{1}{2}\hat{m}''(x_i)\mu_2\right\}\left\{-\frac{\hat{g}'''(0\mid
x_i)}{2\hat{g}''(0\mid x_i)}\right\}.
\end{align*}

\csubsection{Monte Carlo simulations}

\textbf{Example 1:} Generate i.i.d. sample $\{(x_i,y_i),i=1,...,n\}$ from 
\[Y=2\sin(\pi X)+\sigma(X)\epsilon,\]
with $X\sim U(0,1)$, and $\sigma(X)=1+2X$. The error is distributed as $\epsilon\sim0.5N(-1,2.5^2)+0.5N(1,0.5^2)$, such that the mean is 0, the mode is approximately 1 and the median is around 0.67. The proposed modal regression function is Mode$(Y|X)=2\sin(\pi X)+1+2X$, the median regression is Median$(Y|X)=2\sin(\pi X)+0.67+1.34X$, and the traditional mean regression is $\e(Y\mid X)=2\sin(\pi X)$. We consider the following four regression estimates: 1) local linear estimate (LL); 2) local M-estimate (LM); 3) Local median regression estimate (LMD); 4) the proposed local linear modal regression (LLMR). The sample sizes $n=200$, $400$ and $800$ are conducted over 500 repetitions.

Note that the above four regression estimates are targeting different regression functions. That is, LL and LM are targeting the mean regression function, LMD is targeting the median regression function, and LLMR is targeting the modal regression function. In order to compare the performance of different estimates, we will look at their prediction performance based on the coverage probabilities when doing prediction based on the same length of small intervals centered around each estimate. For the error distribution assumed above, the standard deviation is close to $\sigma=2$, therefore, the lengths of intervals considered are 0.1$\sigma$, 0.2$\sigma$, and 0.5$\sigma$. The coverage probabilities are approximated by doing prediction for the 1,000 equally spaced grid points from 0.1 to 0.9, with 500 repetitions.

Table \ref{tab2} contains the average and standard deviation of the estimated coverage probabilities when doing prediction based on the same length of intervals centered around each estimate. From Table \ref{tab2} we can see that LLMR provides the highest coverage probability among the methods considered. In addition, LMD also provides better prediction performance than the mean regression estimates LL and LM, partly due to the skewness of the error distribution.

\begin{table}[htb]
    \centering
\caption{Average (Std) of percentage of coverage with $\sigma=2$.} \vskip 0.05in
\def\arraystretch{1.5}
\small \hspace*{-22.75pt}
\begin{tabular}{c c| c c cc} \hline
Width & Method & n=200 & n=400 & n=800\\
\hline
0.1$\sigma$ & LL&0.030(0.007)&0.029(0.005)&0.028(0.004)\\
& LM&0.047(0.011)&0.047(0.008)&0.047(0.007)\\
& LMD&0.073(0.009)&0.075(0.007)&0.077(0.005)\\
& LLMR&0.081(0.014)&0.086(0.011)&0.090(0.009)\\
\hline
0.2$\sigma$ & LL&0.062(0.014)&0.059(0.011)&0.057(0.008)\\
& LM&0.095(0.021)&0.095(0.017)&0.095(0.013)\\
& LMD&0.145(0.017)&0.149(0.012)&0.153(0.009)\\
& LLMR&0.160(0.026)&0.169(0.021)&0.176(0.017)\\
\hline
0.5$\sigma$ & LL&0.173(0.033)&0.169(0.026)&0.166(0.020)\\
& LM&0.248(0.044)&0.252(0.034)&0.254(0.027)\\
& LMD&0.340(0.031)&0.350(0.020)&0.356(0.014)\\
& LLMR&0.366(0.046)&0.382(0.036)&0.393(0.027)\\
\hline
\end{tabular}
\label{tab2}
\end{table}

\textbf{Example 2:} In this example, we demonstrate the performance of the proposed nonparametric varying coefficient modal regression by the following two models:\\
\textbf{Model 1:} $y=g_0(u)+g_1(u)x_1+g_2(u)x_2+\sigma(u)\epsilon$, where $g_0(u)=\exp(2u-1)$, $g_1(u)=8u(1-u)$, and $g_2(u)=2\sin^2(2\pi u)$.\\
\textbf{Model 2:} $y=g_0(u)+g_1(u)x_1+g_2(u)x_2+\sigma(u)\epsilon$, where $g_0(u)=\sin(2\pi u)$, $g_1(u)=(2u-1)^2+0.5$, and $g_2(u)=\exp(2u-1)-1$.

In both models, $x_1$ and $x_2$ follow a standard normal distribution with correlation coefficient $1/\sqrt{2}$. The index variable $u$ is uniformly distributed on $[0,1]$, and is independent of $(x_1,x_2)$. Similar to the previous example, we consider $\epsilon\sim0.5N(-1,2.5^2)+0.5N(1,0.5^2)$, and $\sigma(u)=1+2u$.

We conduct simulations with sample sizes $n=200$, $400$, and $800$, respectively, with 200 data replications. The bandwidths for LL, LM, and LMD are chosen by cross-validation. 

To compare the coverage probabilities of all methods, we take 30 equally spaced points from 0.1 to 0.9 for $x_1$, $x_2$, and $u$, and do prediction for all of the 9,000 grid points. Tables \ref{tab4} and \ref{tab5} contain the estimated coverage probabilities for Model 1 and Model 2, respectively, based on the same length of small intervals centered around each estimate. From Tables \ref{tab4} and \ref{tab5}, we can see that LLMR provides higher coverage probabilities than all the other three methods, which becomes even more obvious when the sample size increases. In addition, LMD and LM also provide higher coverage probabilities than LL.


\begin{table}[htb]
    \centering
\caption{Average (Std) of percentage of coverage of Model 1, with $\sigma=2$.} \vskip 0.05in
\def\arraystretch{1.5}
\small \hspace*{-22.75pt}
\begin{tabular}{c c| c c cc} \hline
Width & Method & n=200 & n=400 & n=800\\
\hline
0.1$\sigma$ & LL&0.031(0.008)&0.028(0.005)&0.027(0.003)\\
& LM&0.041(0.010)&0.040(0.007)&0.036(0.005)\\
& LMD&0.047(0.010)&0.043(0.007)&0.041(0.005)\\
& LLMR&0.067(0.013)&0.076(0.012)&0.080(0.013)\\
\hline
0.2$\sigma$ & LL&0.062(0.015)&0.058(0.010)&0.056(0.007)\\
& LM&0.084(0.020)&0.081(0.014)&0.075(0.010)\\
& LMD&0.094(0.020)&0.088(0.014)&0.084(0.010)\\
& LLMR&0.133(0.026)&0.150(0.022)&0.158(0.024)\\
\hline
0.5$\sigma$ & LL&0.171(0.037)&0.164(0.025)&0.163(0.018)\\
& LM&0.222(0.044)&0.220(0.031)&0.208(0.024)\\
& LMD&0.243(0.041)&0.233(0.030)&0.226(0.022)\\
& LLMR&0.320(0.049)&0.353(0.038)&0.366(0.043)\\
\hline
\end{tabular}
\label{tab4}
\end{table}

\begin{table}[htb]
    \centering
\caption{Average (Std) of percentage of coverage of Model 2, with $\sigma=2$.} \vskip 0.05in
\def\arraystretch{1.5}
\small \hspace*{-22.75pt}
\begin{tabular}{c c| c c cc} \hline
Width & Method & n=200 & n=400 & n=800\\
\hline
0.1$\sigma$ & LL&0.033(0.008)&0.030(0.006)&0.029(0.004)\\
& LM&0.047(0.011)&0.042(0.008)&0.040(0.006)\\
& LMD&0.052(0.010)&0.050(0.008)&0.049(0.005)\\
& LLMR&0.068(0.014)&0.077(0.012)&0.084(0.012)\\
\hline
0.2$\sigma$ & LL&0.066(0.017)&0.062(0.012)&0.059(0.007)\\
& LM&0.094(0.021)&0.085(0.017)&0.082(0.011)\\
& LMD&0.106(0.019)&0.102(0.016)&0.099(0.010)\\
& LLMR&0.135(0.027)&0.152(0.024)&0.166(0.022)\\
\hline
0.5$\sigma$ & LL&0.184(0.039)&0.177(0.030)&0.171(0.019)\\
& LM&0.248(0.045)&0.230(0.036)&0.225(0.025)\\
& LMD&0.271(0.038)&0.266(0.032)&0.261(0.021)\\
& LLMR&0.323(0.052)&0.355(0.042)&0.378(0.038)\\
\hline
\end{tabular}
\label{tab5}
\end{table}

\csubsection{Health Care Expenditure data.}
We illustrate the proposed methodology by an analysis of the health care expenditure data (Cohen, 2003; Natarajan et al., 2008). The data set comes from the Medical Expenditure Panel Survey (MEPS) for the year 2002, which was conducted by the United States National Center for Health
Statistics, Centers for Disease Control and Prevention. The survey was designed to produce
national and regional estimates of the health care use, expenditures, sources of payment and
insurance coverage of the US civilian non-institutionalized population. Medical cost data are typically highly skewed to the
right, in that a small percentage of subjects sustain extremely high costs compared to
other subjects.

We randomly select 500 patients within one ``primary sampling units'' (PSUs) as our data example. The outcome of interest is ``total health care expenditures in the year 2002''. The covariates of interest are race (1 if white; 0 if otherwise), smoke (1 if a current
smoker; 0 if otherwise), pov (1 if above the poverty line; 0 if at or below the poverty line), insur
(1 if the patient has health insurance; 0 if otherwise), phealth (1 if good perceived health status;
0 if otherwise) and meds (1 if the patient needs prescription medication; 0 if otherwise), and we take $u$=age. We fit the data by LL, LM, LMD, and LLMR.

With $10\%, 30\%, 50\%,$ and $90\%$ as the levels of confidence, Table \ref{tab6} reports the average widths and percentage of coverage of the prediction intervals. The confidence interval of LLMR is constructed based on the similar method suggested by Yao and Li (2014), which could make use of the skewness of the error distribution assumed by LLMR. The coverage probability is measured by leave-one-out cross validation. From Table \ref{tab6}, we can see that the actual coverage rates are very close to the nominal confidence levels for all methods. The average widths of LMD and LLMR are shorter compared to LL and LM, and LLMR is superior for higher confidence levels.

To evaluate the prediction performance of the methods, we apply $d$-fold cross-validation and Monte-Carlo cross-validation (MCCV) to the data, and the median and standard deviation of the median of squared prediction errors (MSPE) are reported in Table \ref{tab7}. The medians of LLMR are much smaller than the other three methods, indicating that LLMR provides the best point prediction followed by LMD and LM.

\begin{table}[htb]
    \centering
\caption{Average widths (percentage of coverage) of the prediction intervals for health care expenditure data.} \vskip 0.05in
\def\arraystretch{1.5}
\small \hspace*{-22.75pt}
\begin{tabular}{c | c c c cc} \hline
Method & 10\%&30\%&50\%&90\%\\
\hline
LL&0.095(0.098)&0.255(0.290)&0.438(0.470)&1.904(0.886)\\
LM&0.080(0.104)&0.215(0.272)&0.394(0.486)&1.897(0.888)\\
LMD&0.053(0.094)&0.163(0.252)&0.381(0.438)&1.758(0.876)\\
LLMR&0.057(0.096)&0.172(0.276)&0.334(0.470)&1.719(0.902)\\
\hline
\end{tabular}
\label{tab6}
\end{table}

\begin{table}[htb]
    \centering
\caption{Median (Std) of MSPE for health care expenditure data.} \vskip 0.05in
\def\arraystretch{1.5}
\small \hspace*{-22.75pt}
\begin{tabular}{c | c c c cc} \hline
Method & 5-fold CV & 10-fold CV & MCCV d=50 & MCCV d=100\\
\hline
LL&0.105(0.020)&0.102(0.029)&0.106(0.033)&0.108(0.023)\\
LM&0.079(0.018)&0.072(0.027)&0.079(0.023)&0.079(0.015)\\
LMD&0.042(0.010)&0.042(0.018)&0.041(0.020)&0.043(0.013)\\
LLMR&0.021(0.009)&0.021(0.010)&0.020(0.020)&0.025(0.016)\\
\hline
\end{tabular}
\label{tab7}
\end{table}

\csection{\sc Concluding Remarks}
In this article, we proposed a nonparametric modal regression and a nonparametric varying coefficient modal regression. Compared to traditional mean regression models, the new nonparametric modal regression models are more robust and have better prediction performance. We demonstrated such superior performance through a simulation study and a health care expenditure data.

Choosing the bandwidths has long been a difficult problem for nonparametric and semiparametric models. In this paper, we propose to use the plug-in method to choose the bandwidths based on the found asymptotic optimal bandwidths. One might also use a sequence of bandwidths as suggested by Kemp and Santos Silva (2012) to reveal some more interesting features of modal regression. In addition, it is also interesting to know how to adapt the traditional cross validation technique to choose the bandwidth for nonparametric modal regression.

The development of modal regression is still in its initial stage. We believe that modal regression could be a good alternative to the mean regression and median regression and there are still much work to be done in the future. Much of the development for mean regression and median regression could have similar development for modal regression.


\scsection{\sc Appendix}

\renewcommand{\theequation}{A.\arabic{equation}}
\setcounter{equation}{0}
\renewcommand{\thelemma}{A.\arabic{lemma}}

\label{sec:proof} 
\renewcommand{\theequation}{A.\arabic{equation}}
\setcounter{equation}{0}
\renewcommand{\thesection}{A}

\label{sec:proof} The conditions used by the theorems are listed below. They are not the weakest possible conditions, but
they are imposed to facilitate the proofs.

\noindent{\bf Technical Conditions:}
\begin{description}
\item{(A1)} The $m(x)$ has continuous ${(p+1)}^{th}$ derivative at
the point $x_0$.
\item{(A2)} $g'(0\mid x)=0, g''(0\mid x)<0$, $g^{(v)}(t\mid x)$ is
bounded in a neighbor of $x_0$ and has continuous first derivative
at the point $x_0$ as a function of $x$, for $v=0,\ldots,4$.
\item{(A3)} The $f(x)$ is bounded and has
continuous first derivative at the point $x_0$ and $f(x_0)>0$.
\item{(A4)} $K(\cdot)$ is a symmetric (about 0) probability
density with compact support $[-1,1]$.
 \item{(A5)} $g_j(x)$ has continuous $2^{nd}$ derivative at
the point $x_0$, $j=1,...,p$.
\item{(A6)} $q'(0\mid \bx,u)=0, q''(0\mid \bx,u)<0$, $q^{(v)}(t\mid \bx,u)$ is
bounded in a neighbor of $(\bx_0,u_0)$ and has continuous first derivative
at the point $(\bx_0,u_0)$ as a function of $(\bx,u)$, for $v=0,\ldots,4$.
\item{(A7)} The $f(u)$ is bounded and has
continuous first derivative at the point $u_0$ and $f(u_0)>0$.
\end{description}
\bigskip

Denote
$X^*_i=\left\{1,(X_i-x_0)/h_1,\ldots,(X_i-x_0)^p/h_1^p\right\}^T$,
$H=\text{diag}\{1,h_1,\ldots,h_1^p\}$,
$\boldtheta=(\beta_0,\beta_1,\ldots,\beta_p)^T,$
$\boldtheta^*=H\boldtheta$,
$R(X_i)=m(X_i)-\sum_{j=0}^p\beta_j(X_i-x_0)^j$,
 and $K_i=K_{h_1}(x_i-x_0)$, where $\beta_j=m^{(j)}(x_0)/j!,
 j=0,1,\ldots,p$.

\medskip

\noindent{\it Proof of Theorem~\ref{modconsis}.} Denote
$\alpha_n=(nh_1h_2^3)^{-1/2}+h_1^{p+1}+h_2^2$. It is sufficient to
show that for any given $\eta>0$, there exists a large constant
$c$ such that
\begin{equation}
P\{\sup_{|\mu\|=c} \ell(\boldtheta^*+\alpha_n \mu) <
\ell(\boldtheta^*)\}\ge 1-\eta, \label{a3}
\end{equation}
where $\ell(\boldtheta)$ is defined in (\ref{lpmod}).

 By using Taylor expansion, it follows that
\begin{align*}
\ell(\boldtheta^*+\alpha_n\mu)-\ell(\boldtheta^*)=
&\frac{1}{n}\sum_{i=1}^nK_i\left\{-\phih'(\epsilon_i+R(X_i))\alpha_n\mu^TX_i^*+\frac{1}{2}\phih''
(\epsilon_i+R(X_i))\alpha_n^2(\mu^TX_i^*)^2\right.\notag\\
&\left.-\frac{1}{6}\phih'''(z_i)\alpha_n^3(\mu^TX_i^*)^3)
\right\}\notag\\
\triangleq &I_1+I_2+I_3,
\end{align*}
where 
$z_i$ is between $\epsilon_i+R(X_i)$ and
$\epsilon_i+R(X_i)+\alpha_n\mu^TX_i^*$. Note that
\[\phih'(t)=-\frac{t}{h_2^3}\phi\left(\frac{t}{h_2}\right), \phih''(t)=\frac{1}{h_2^{3}}
\left(\frac{t^2}{h_2^2}-1\right)\phi\left(\frac{t}{h_2}\right),
\text{ and }
\phih'''(t)=\frac{1}{h_2^{4}}\left\{\frac{3t}{h_2}-\left(\frac{t}{h_2}\right)^3\right\}\phi\left(\frac{t}{h_2}\right).\]

If $a_n(x)=o_p(h_2)$, and $g^{(v)}(t\mid x)$ is bounded in a
neighbor of $x_0$, we have
\begin{align*}
\e\{\phih'(\epsilon+a_n(x))\mid X=x\}&=-h_2^{-1}\int t\phi(t)g(th-a_n(x)|x)dt\notag\\
&=-\left\{\frac{g'''(0\mid
x)}{2}h_2^2+g''(0|x)a_n(x)\right\}\{1+o_p(1)\}. 
\end{align*}
If $h_1^{p+1}/h_2\rightarrow 0$, by directly calculating the mean
and variance, we obtain
\begin{align}\label{meanvar}
\e(I_1)
&=\alpha_n\mu^T\left\{\frac{g'''(0\mid x_0)}{2}f(x_0)\breve{c}_ph_2^2-
g''(0|x_0)c_pf(x_0)\frac{m^{(p+1)}(x_0)}{(p+1)!}h_1^{p+1}\right\}\notag\\
&=O\left\{\alpha_nc\left(h_2^2+h_1^{p+1}\right)\right\},\notag\\
\var(I_1)
&=n^{-1}\alpha_n^2\mu^T\left\{g(0|x_0)f(x_0)\nu_0S^*h_2^{-3}h_1^{-1}\right\}\mu\notag\\
&=O(\alpha_n^2(nh_1h_2^3)^{-1}c^2),
\end{align}
where $\breve{c}_p=(\mu_0,\mu_1,\ldots,\mu_p)^T$. Hence $I_1=O\left\{\alpha_nc\left(h_2^2+h_1^{p+1}\right)\right\}+\alpha_ncO_p((nh_1^{-1}h_2^{-3})^{-1/2})=O_p(c\alpha_n^2).$
If $nh_2^5h_1\rightarrow \infty$, similar to (\ref{meanvar}), we
can prove
\begin{align}
I_2&=\frac{1}{n}\sum_{i=1}^n\left\{\frac{1}{2}K_i\phih''
(\epsilon_i+R(X_i))\alpha_n^2\mu^TX_i^*{X_i^*}^T\mu\right\}=\alpha_n^2g''(0|x_0)f(x_0)\mu^TS\mu(1+o_p(1)),\notag\\
I_3&=\frac{1}{n}\sum_{i=1}^n\left\{-\frac{1}{6}K_i\phih'''(z_i)\alpha_n^3(\mu^TX_i^*)^3\right\}=o_p(\alpha_n^2).\label{i3}
\end{align}

Noticing that $S$ is a positive matrix, $\|\mu\|=c$, and
$g''(0|x_0)<0$, we can choose $c$ large enough such that $I_2$
dominates both $I_1$ and $I_3$ with probability at least $1-\eta$.
Thus (\ref{a3}) holds.
Therefore, with probability approaching 1 (wpa1), there exists a
local maximizer $\hat{\boldtheta}^*$ such that
$||\hat{\boldtheta}^*-\boldtheta^*||\le\alpha_nc$. Based on the
definition of $\boldtheta^*$, we can get, wpa1,
$\left|h_1^v\left\{\hat{m}_v(x_0)-m^{(v)}(x_0)\right\}\right|=O_p\left\{(nh_1h_2^3)^{-1/2}+h_1^{p+1}+h_2^2\right\}$.
\qed

Define
\begin{equation}
W_n=\sum_{i=1}^nX_i^*K_i\phih'(\epsilon_i). \label{defwn}
\end{equation}
We have the following asymptotic representation.

\begin{lem}\label{lem:rep}
Under conditions (A1)|(A4), it follows that
\begin{equation}
\hat{\boldtheta}^*-\boldtheta^*
=h_1^{p+1}\frac{m^{(p+1)}(x_0)}{(p+1)!}S^{-1}c_p(1+o_p(1))+\frac{S^{-1}W_n}{nF(x_0,h_2)f(x_0)}(1+o_p(1)).
\label{eq:rep}
\end{equation}
\end{lem}

{\bf Proof.} Let $\hat{\gamma}_i=R(X_i)-\sum_{j=0}^p(\hat{\beta}_j-\beta_j)(X_i-x_0)^j=R(X_i)-(\hat{\boldtheta}^*-\boldtheta^*)^TX^*$, then $Y_i-\sum_{j=0}^p\hat{\beta}_j(X_i-x_0)^j=\epsilon_i+\hat{\gamma}_i$. The solution $\hat{\boldtheta}^*$ satisfies the equation
\begin{equation} \label{eq:score}
\sum_{i=1}^nX_i^*K_i\phih'(\epsilon_i+\hat{\gamma}_i)=\sum_{i=1}^nX_i^*K_i\left\{\phih'(\epsilon_i)
+\phih''(\epsilon_i)\hat{\gamma}_i+\frac{1}{2}\phih'''(\epsilon^*)\hat{\gamma}^2\right\}=0,
\end{equation}
where $\epsilon^*$ is between $\epsilon_i$ and
$\epsilon_i+\hat{\gamma}_i$. Note that the second term on the left
hand side of (\ref{eq:score}) is
\begin{equation}
\label{ab}
\sum_{i=1}^nK_i\phih''(\epsilon_i)R(X_i)X_i^*-\sum_{i=1}^nK_i\phih''(\epsilon_i)
X_i^*{X_i^*}'(\hat{\boldtheta}^*-\boldtheta^*) \triangleq   J_1+J_2.
\end{equation}
From the proof of (\ref{meanvar}), we have
\begin{align*}
J_1&=nh_1^{p+1}g''(0\mid x_0)f(x_0)
c_p\frac{m^{(p+1)}(x_0)}{(p+1)!}+o_p(nh_1^{p+1}),
\end{align*}
and
\[J_2=-ng''(0\mid x_0)f(x_0)S(1+o_p(1))
(\hat{\boldtheta}^*-\boldtheta^*).\]

From Theorem \ref{modconsis}, we know
$||\hat{\boldtheta}^*-\boldtheta^*||=O_p\{h_1^{p+1}+h_2^2+(nh_1h_2^3)^{-1/2}\}$,
hence
 \begin{align}
\sup_{i:|X_i-x_0|/h\le 1}|\hat{\gamma}_i|&\le
\sup_{i:|X_i-x_0|/h\le 1} |R(X_i)|+(\hat{\boldtheta}^*-\boldtheta^*)^TX^*\notag\\
&=O_p(h_1^{p+1}+||\hat{\boldtheta}^*-\boldtheta^*||)=O_p(||\hat{\boldtheta}^*-\boldtheta^*||).
\label{gamorder}
\end{align}
($m^{(p+1)}(x)$ is bounded)  
Also, we have
\begin{align*}
\label{eq:z2} \e\left\{K_i(X_i-x_0)^j/h_1^j\right\}&=
\int\frac{1}{h_1}K\left(\frac{x-x_0}{h_1}\right)\left(\frac{x-x_0}{h}\right)^jf(x)dx \nonumber\\
&=\mu_jf(x_0)+o(1).
\end{align*}
Based on (\ref{i3}), (\ref{gamorder}), and
$||\hat{\boldtheta}^*-\boldtheta^*||=O_p(\alpha_n)$,
\begin{align}
\frac{1}{n}\left\{\sum_{i=1}^nK_i\hat{\gamma}^2_i\phih'''(\epsilon^*)(X_i-x_0)^j/h_1^j\right\}
&=\frac{1}{n}\left\{\sum_{i=1}^nK_i\alpha_n^2\phih'''(\epsilon^*)(X_i-x_0)^j/h_1^j\right\}= o_p(||\alpha_n||)\nonumber.
\end{align}
Hence for the third term on the left-hand side of
(\ref{eq:score}),
\[\sum_{i=1}^nK_i\hat{\gamma}^2_iX_i^*\phih'''(\epsilon^*)=o_p(n\alpha_n)=o_p(J_2).\]

Then, it follows from (\ref{defwn}) and (\ref{eq:score}) that
\begin{align*}
\hat{\boldtheta}^*-\boldtheta^*
=&h_1^{p+1}\frac{m^{(p+1)}(x_0)}{(p+1)!}S^{-1}c_p(1+o_p(1))+\frac{S^{-1}W_n}{ng''(0\mid
x_0)f(x_0)}(1+o_p(1)).
\end{align*}
\qed
\bigskip

\noindent{\it Proof of Theorem~\ref{modasybv}}. Based on
(\ref{defwn}) and the condition (A6), we can easily get
\[\e\left(\frac{1}{n}W_n\right)=\e\left\{X_i^*K_i\phih'(\epsilon_i)\right\}=-\frac{g'''(0\mid x_0)}{2}f(x_0)c_p^*h_2^2+o_p(h_2^2).\]

Similar to the proof of (\ref{meanvar}), we have
\begin{align*}
E\left\{K_i^2\phih'(\epsilon_i)^2(X_i-x_0)^j/h_1^j\right\}
=(h_1h_2^3)^{-1}\nu_j\tilde{\nu}g(0\mid x_0)f(x_0)\{1+o(1)\}.
\end{align*}
where $\tilde{\nu}=\int\phi^2(t)t^2dt$. So
\begin{equation}
\cov(W_n/n)=(nh_1h_2^3)^{-1}\tilde{\nu}g(0\mid
x_0)f(x_0)S^*(1+o(1)). \label{eq:covw}
\end{equation}
Based on the result (\ref{eq:rep}), the asymptotic bias $b_v(x_0)$
and variance of $\hat{m}_v(x_0)$ are naturally given by
\[
b_v(x_0)=e_{v+1}^TS^{-1}\left\{h_1^{p+1-v}\frac{v!}{(p+1)!}m^{(p+1)}(x_0)c_p-\frac{g'''(0\mid
x_0)v!h_2^2}{2g''(0\mid x_0)h_1^v}c^*_p\right\}(1+o(1))\] and
\[\var\{\hat{m}_v(x_0)\}=\frac{v!^2g(0\mid x_0)\tilde{\nu}}{nh_2^3h_1^{1+2v}f(x_0)g''(0\mid x_0)^2}e_{v+1}^TS^{-1}S^{*}S^{-1}
e_{v+1}(1+o(1)).
\]

Noting that $\mu_j=0$ for odd $j$, by some simple calculation, we
can know the $(v+1)^{th}$ element of $S^{-1}c_p$ is zero for $p-v$
even. So we need higher order expansion of asymptotic bias for
$p-v$ even. Following the similar arguments of Theorem
\ref{modconsis}, if $nh_1^3h_2^5\rightarrow \infty$ (make the root
of variance order less than bias order), we can prove
\begin{align*}
J_1&=nh_1^{p+1}\left[\Gamma(x_0)
c_p\frac{m^{(p+1)}(x_0)}{(p+1)!}+h_1\tilde{c}_p\left\{\Gamma'(x_0)\frac{m^{(p+1)}(x_0)}{(p+1)!}+\Gamma(x_0)
\frac{m^{(p+2)}(x_0)}{(p+2)!}\right\}\right]\{1+o_p(1)\}, \nonumber \\
J_2&=-n\left\{\Gamma(x_0)S+h_1\tilde{S}\Gamma'(x_0)\right\}\{1+o_p(1)\}
(\hat{\boldtheta}^*-\boldtheta^*),
\end{align*}
where $J_1$ and $J_2$ are defined in (\ref{ab}) and
$\Gamma(x)=g''(0\mid x)f(x)$.

Then, it follows from (\ref{eq:score}) that
\begin{align*}
\hat{\boldtheta}-\boldtheta
&=h_1^{p+1}\left\{\frac{m^{(p+1)}(x_0)}{(p+1)!}S^{-1}c_p+h_1b^*(x_0)\right\}(1+o_p(1))
+\frac{S^{-1}W_n}{n\Gamma(x_0)}(1+o_p(1)),
\end{align*}
where
\begin{align*}
b^*(x_0)=&\Gamma^{-1}(x_0)S^{-1}\tilde{c}_p\left\{\Gamma'(x_0)\frac{m^{(p+1)}(x_0)}{(p+1)!}
+\Gamma(x_0)\frac{m^{(p+2)}(x_0)}{(p+2)!}\right\}\\
&-\Gamma^{-1}(x_0)\Gamma'(x_0)\frac{m^{(p+1)}(x_0)}{(p+1)!}S^{-1}\tilde{S}S^{-1}c_p\;.
\end{align*}

For $p-v$ even, since the $(v+1)^{th}$ element of $S^{-1}c_p$ and
$S^{-1}\tilde{S}S^{-1}c_p$ are zeros, the asymptotic bias $b_v(x_0)$
of $\hat{m}_v(x_0)$ are naturally given by
\begin{align*}
b_v(x_0)=&e_{v+1}^TS^{-1}\left[\tilde{c}_p\frac{h_1^{p+2-v}v!}{(p+2)!}\left\{m^{(p+2)}(x_0)+(p+2)m^{(p+1)}(x_0)
\frac{\Gamma'(x_0)}{\Gamma(x_0)}\right\}\right.\\
&\ \ \left.-\frac{g'''(0\mid x_0)v!h_2^2}{2g''(0\mid
x_0)h_1^v}c^*_p\right]\left\{1+o(1)\right\}.
\end{align*}
\qed
\bigskip

\noindent{\it Proof of Theorem \ref{modnorm}}.

It is sufficient to show that
\begin{equation}
W_n^*\equiv\sqrt{h_1h_2^3/n}W_n\wln N(0, D), \label{asynormwn}
\end{equation}
where $D=\tilde{\nu}g(0\mid x_0)f(x_0)S^*,$ because using
Slutsky's theorem , it follows from (\ref{eq:rep}),
(\ref{asynormwn}), and Theorem \ref{modasybv} that
\[
\frac{\hat{m}_v(x_0)-m^{(v)}(x_0)-b_v(x_0)}{\sqrt{\var\{\hat{m}_v(x_0)\}}}
\wln N(0,1).
\]

Next we show (\ref{asynormwn}). For any unit vector $d\in
\mathbb{R}^{p+1}$, we prove
\[\{d^T\cov(W_n^*)d\}^{-\frac{1}{2}}\{d^TW_n^*-d^TE(W_n^*)\}\wln N(0,1).\]

Let
\[\xi_i=\sqrt{h_1h_2^3/n}K_i\phih'(\epsilon_i)d^TX_i^*.\]
Then $d^TW_n^*=\sum_{i=1}^n\xi_i$. We check the Lyapunov's
condition.
Based on (\ref{eq:covw}), we can get
$\cov(W_n^*)=\tilde{\nu}g(0\mid x_0)f(x_0)S^*(1+o(1))$ and
$\var(d^TW_n^*d)=d^T\cov(W_n^*)d=\tilde{\nu}g(0\mid
x_0)f(x_0)d^TS^*d(1+o(1))$. So we only need to prove
$nE|\xi_1|^3\rightarrow 0$. Noticing that $(d'X_i)^2\le
||d||^2||X_i||^2, \phi'(\cdot)$ is bounded, and $K(\cdot)$ has
compact support,
\begin{align*}
nE|\xi|^3&\le
O(nn^{-3/2}h_1^{3/2}h_2^{9/2})\sum_{j=0}^pE\left|K_1^3\phih'(\epsilon_1)^3\left(\frac{X_1-x_0}{h_1}
\right)^{3j}\right|
\rightarrow 0.
\end{align*}
So the asymptotic normality for $W_n^*$ holds with covariance
matrix $\tilde{\nu}g(0\mid x_0)f(x_0)S^*$.
\qed

\bigskip
\noindent{\it Proof of Theorem~\ref{modasyvarcoef}}.
The proof is similar to Theorem~\ref{modasybv} and Theorem \ref{modnorm}. Here, we provide a sketch of the proof. Let $\btheta_0$ be the true value of $\btheta$. Note that when $u_i$ is close to $u_0$, we have
\begin{align*}
y_i-\tilde{x}_i^T\btheta_0&=\epsilon_i+m(\bx_i,u_i)-\sum_{j=1}^p\{g_j(u_0)+g_j'  (u_0)(u_i-u_0)\}x_{ij}\\
&\triangleq \epsilon_i+R(\bx_i,u_i)=\epsilon_i+\frac{1}{2}\sum_{j=1}^pg_j''(u_0)(u_i-u_0)^2x_{ij}(1+o_p(1)).
\end{align*}
For simplicity of notations, we denote $R(\bx_i,u_i)$ by $R_i$, and $q(\epsilon\mid \bx_i,u_i)$ by $q_i(\epsilon)$. Then the objective function is
\begin{equation*}
\label{eq:defloclh}
\ell(\btheta_0)=\frac{1}{n}\sum_{i=1}^nK_{h_1}(u_i-u_0)\phi_{h_2}(y_i-\tilde{\bx}_i^T\btheta_0)
=\frac{1}{n}\sum_{i=1}^nK_{h_1}(u_i-u_0)\phi_{h_2}(\epsilon_i+R_i).
\end{equation*}
Note that
\begin{align*}
\e\left[\phi_{h_2}''(\epsilon_i+R_i)\mid \bx_i,u_i\right]
&=q_i''(0)(1+o_p(1)),\\
\e\left[\phi_{h_2}'(\epsilon_i+R_i)\mid \bx_i,u_i\right]
&=-\left[q_i'''(0)h_2^2/2-q_i''(0)R_i\right](1+o_p(1)),
\end{align*}
and
\begin{align*}
\ell'(\btheta_0)
&=\frac{1}{n}\sum_{i=1}^nK_{h_1}(u_i-u_0)\phi_{h_2}'(e_i+R_i)\tilde{\bx}_i.
\end{align*}
Let $\alpha_j(u_i)=\e\{\bx_ix_{ij}q_i''(0)\mid u_i\}$ and $\beta(u_i)=\e\{\bx_i q_i'''(0)\mid u_i\}$, then we have
\begin{align*}
\e\{\ell'(\btheta_0)\}
&=\left\{-\frac{h_2^2}{2}f(u_0)\binom{1}{\mu_1}\otimes\beta(u_0)+\frac{h_1^2}{2}\sum_{j=1}^pg_j''(u_0)f(u_0)\binom{\mu_2}{\mu_3}\otimes\alpha_j(u_0)\right\}(1+o_p(1)).
\end{align*}
Note that
\begin{align*}
\ell''(\btheta_0)
&=\frac{1}{n}\sum_{i=1}^nK_{h_1}(u_i-u_0)\phi_{h_2}''(e_i+R_i)\tilde{\bx}_i\tilde{\bx}_i^T,
\end{align*}
then we have
\begin{align*}
\e\{\ell''(\btheta_0)\}
&=f(u_0)\bigl(\begin{smallmatrix}1 &\mu_1 \\ \mu_1 &\mu_2 \end{smallmatrix}\bigr)\otimes\Delta(u_0)(1+o_p(1)).
\end{align*}
In addition, since
\begin{align*}
\e\{\phi'_{h_2}(\epsilon_i+R_i)^2\mid u_i,\bx_i\}&=h_2^{-3}\int t^2\phi^2(t)q_i(th-R_i)dt
=h_2^{-3}q_i(0)\tilde{\nu}(1+o_p(1)),
\end{align*}
then,
\begin{align}
\var\{\ell'(\btheta_0)\}&=\frac{\tilde{\nu}}{nh_2^3}\e[K_{h_1}^2(u_i-u_0)q_i(0)\tilde{\bx}_i\tilde{\bx}_i^T](1+o_p(1))\notag\\
&=\frac{\tilde{\nu}}{nh_2^3h_1}f(u_0)\bigl(\begin{smallmatrix}\nu_0 &\nu_1 \\ \nu_1 &\nu_2 \end{smallmatrix}\bigr)\otimes\tilde{\Delta}(u_0)(1+o_p(1)).
\label{eq:asyvarl1}
\end{align}

Therefore,
\begin{equation*}
Bias\left\{\binom{\hat{\textbf{g}}(u_0)}{\hat{\textbf{g}}'(u_0)}\right\}=\frac{1}{2}\Delta^{-1}(u_0)\otimes\binom{\mu_2h_1^2\sum_{j=1}^pg_j''(u_0)\alpha_j(u_0)-h_2^2\beta(u_0)}
{\mu_3h_1^2\sum_{j=1}^pg_j''(u_0)\alpha_j(u_0)-\mu_1h_2^2\beta(u_0)}\{1+o_p(1)\},
\end{equation*}
and
\begin{equation*}
\cov\left\{\binom{\hat{\textbf{g}}(u_0)}{\hat{\textbf{g}}'(u_0)}\right\}=\frac{\tilde{\nu}}{nh_1h_2^3f(u_0)}\bigl(\begin{smallmatrix}\nu_0 &\frac{\nu_1}{\mu_2} \\ \frac{\nu_1}{\mu_2} &\frac{\nu_2}{\mu^2_2} \end{smallmatrix}\bigr)\otimes\Delta^{-1}(u_0)\tilde{\Delta}(u_0)\Delta^{-1}(u_0)(1+o_p(1)).
\end{equation*}
Specifically,
\begin{equation*}
Bias\left\{\hat{\textbf{g}}(u_0)\right\}=\frac{1}{2}\Delta^{-1}(u_0)\left\{\mu_2h_1^2\sum_{j=1}^pg_j''(u_0)\alpha_j(u_0)-h_2^2\beta(u_0)\right\}\left\{1+o_p(1)\right\},
\end{equation*}
and the asymptotic variance is
\begin{equation*}
\cov\left\{\hat{\textbf{g}}(u_0)\right\}=\frac{\tilde{\nu}\nu_0}{nh_1h_2^3f(u_0)}\Delta^{-1}(u_0)\tilde{\Delta}(u_0)\Delta^{-1}(u_0)\left\{1+o_p(1)\right\}.
\end{equation*}
\qed

\bigskip
\noindent{\it Proof of Theorem~\ref{modasynorm}.}
It is sufficient to show that
\begin{equation}
T_n=\sqrt{nh_1h_2^3}\ell'(\btheta_0)\wln N(0, T)
\label{eq:asyt}
\end{equation}
where $T=\tilde{\nu}f(u_0)\bigl(\begin{smallmatrix}\nu_0 &\nu_1 \\ \nu_1 &\nu_2 \end{smallmatrix}\bigr)\otimes\tilde{\Delta}(u_0)$, then by Slutsky's theorem and Theorem \ref{modasyvarcoef}, we can obtain
\[[\cov\{\hat{\textbf{g}}(u_0)\}]^{-1/2}[\hat{\textbf{g}}(u_0)-\textbf{g}_0(u_0)-\text{bias}\{\hat{\textbf{g}}(u_0)\}]\wln N(0,I).\]
To show (\ref{eq:asyt}), we prove that for any unit vector $d\in
\mathbb{R}^{p+1}$,
\[\{d^T\cov(T_n)d\}^{-\frac{1}{2}}\{d^TT_n-d^TE(T_n)\}\wln N(0,1).\]

By (\ref{eq:asyvarl1}), $\text{Cov}(T_n)=\tilde{\nu}f(u_0)\bigl(\begin{smallmatrix}\nu_0 &\nu_1 \\ \nu_1 &\nu_2 \end{smallmatrix}\bigr)\otimes\tilde{\Delta}(u_0)(1+o_p(1))$, and $\text{Var}(d^TT_nd)=d^T\text{Cov}(T_n)d=\tilde{\nu}f(u_0)d^T\bigl(\begin{smallmatrix}\nu_0 &\nu_1 \\ \nu_1 &\nu_2 \end{smallmatrix}\bigr)\otimes\tilde{\Delta}(u_0)d(1+o_p(1))$. Let $\xi_i=\sqrt{nh_1h_2^3}K_{h_1}(u_i-u_0)\phi'_{h_2}(y_i-\tilde{\bx}_i^T\btheta_0)d^T\tilde{\bx}_i$. Similar to the proof of Theorem \ref{modnorm}, we can show that $E|\xi|^3\rightarrow 0$, and so the asymptotic normality for $T_n$ holds with covariance matrix $T$.
\qed



\newpage

\begin{thebibliography}{99}
\frenchspacing



\bibitem[Chaudhuri and Marron(1999)]{CM99}
Chaudhuri, P. and Marron, J. S. (1999). Sizer for exploration of
structures in curves. \jasa, 94, 807-823.

\bibitem[Chiang et al. (2001)]{chiang2001}
Chiang, C-T., Rice, J. A. and Wu, C. O. (2001). Smoothing spline estimation for varying
coefficient models with repeatedly measured dependent variable. \emph{Journal of the
American Statistical Association}, 96, 605-619.

\bibitem[cohen(2003)]{co2003}
Cohen, S.B. (2003). Design strategies and innovations in the medical expenditure panel
survey. \emph{Medical care}, 41(7):III.




\bibitem[Fan and Gijbels(1996)]{fan96}
Fan, J. and Gijbels, I. (1996). \emph{Local Polynomial Modelling
and Its Applications}. Chapman and Hall, London.


\bibitem[Fan and Zhang (1999)]{fan1999}
Fan, J. and Zhang, W. (1999). Statistical estimation in varying coefficient models. \emph{Annals
of Statistics}, 27, 1491-1518.

\bibitem[Fan and Zhang (2000)]{fan2000}
Fan, J. and Zhang, J. T. (2000). Two-step estimation of functional linear models with
applications to longitudinal data. \emph{Journal of the Royal Statistical Society: Series B}, 62,
303-322.


\bibitem[Friedman and Fisher(1999)]{FF99}
Friedman J. H. and Fisher, N. I. (1999). Bump hunting in
high-dimensional data. \emph{Statistics and Computing}, 9, 123-143.





\bibitem[Hall, Minnotte, and Zhang(2004)]{HMZ04}
Hall, P., Minnotte, M. C., and Zhang, C. (2004). Bump hunting with
non-Gaussian kernels. \emph{Annals of Statistics}, 32, 2124-2141.











\bibitem[Hastie(1993)]{hastie93}
Hastie, T.J. and Tibshirani, R.J. (1993). Varying-coefficient models (with discussion). \emph{Journal of the Royal Statistical Society: Series B}, 55, 757-796.

\bibitem[Hoover(1998)]{hoover98}
Hoover, D. R., Rice, J. A., Wu, C. O. and Yang, L. P. (1998). Nonparametric smoothing
estimates of time-varying coeffcient models with longitudinal data. \emph{Biometrika}, 85,
809-822.

\bibitem[Huang et al. (2002)]{huang2002}
Huang, J. Z., Wu, C. O. and Zhou, L. (2002). Varying-coefficient models and basis
function approximations for the analysis of repeated measurements. \emph{Biometrika}, 89,
111-128.





\bibitem[Kemp and Santos Silva (2012)]{kemp12}
Kemp, G. C. R. and Santos Silva, J. M. C. (2012). Regression towards
the mode. \emph{Journal of Economics}, 170, 92-101.


\bibitem[Lee(1989)]{lee89}
Lee, M. J. (1989). Mode Regression. \emph{Journal of Econometrics},
42, 337-349.

\bibitem[Lee(1992)]{lee92}
Lee, M. J. (1992). Mode Regression. \emph{Journal of Econometrics},
57, 1-19.

\bibitem[Lee(1998)]{lee98}
Lee, M.J. and Kim, H.J. (1998). Semiparametric econometric estimators for a truncated regression model: a review with an extension. \emph{Statistica Neerlandica}, 52, 200-225.

\bibitem[Muller and Sawitzki(1991)]{MS91}
Muller, D. W. and Sawitzki, G. (1991). Excess mass estimates and
tests for multimodality. \jasa, 86, 738-746.

\bibitem[Nadaraya(1964)]{nadaraya64}
Nadaraya, E. A. (1964). On estimating regression. \emph{Theory of
Probability Applied,} 10, 186-190.

\bibitem[Natarajan et al (2008)]{nata2008}
Natarajan, S., Lipsitz, S.R., Fitzmaurice, G., Moore, C.G. and
Gonin, R. (2008). Variance estimation in complex survey sampling for generalized linear
models. \emph{Journal of the Royal Statistical Society: Series C (Applied Statistics)}, 57(1), 75-87.


\bibitem[Ray and Lindsay(2005)]{RL05}
Ray, S. and Lindsay, B. G. (2005). The topography of multivariate
normal mixtures. \ann, 33, 2042-2065.



\bibitem[Scott(1992)]{Scott92}
Scott, D. W. (1992). \emph{Multivariate Density Estimation: Theory,
Practice and Visualization.} New York: Wiley.

\bibitem[Watson(1964)]{watson64}
Watson, G. S. (1964). Smooth regression analysis. \emph{Sankhya},
Ser. A, 26, 359-372.

\bibitem[Wu et al. (1998)]{wu98}
Wu, C. O., Chiang, C. T. and Hoover, D. R. (1998). Asymptotoc confidence regions for
kernel smoothing of a aarying coefficient model with longitudinal data. \emph{Journal of
the American Statistical Association}, 93, 1388-1402.

\bibitem[Yao and Li (2014)]{yao2014}
Yao, W. and Li, L. (2014). A new regression model: modal linear regression. \emph{Scandinavian Journal of Statistics}, 1-16.

\bibitem[Yao and Lindsay(2009)]{yao09}
Yao, W. and Lindsay, B. G. (2009). Bayesian mixture labelling by
highest posterior density. \emph{Journal of American Statistical
Association}, 104, 758-767.

\bibitem[Yao et al.(2012)]{yao12}
Yao, W., Lindsay, B. G., and Li, R. (2012). Local
modal regression. \emph{Journal of Nonparametric Statistics,} 24, 647-663.


\end{thebibliography}
\end{document}